\documentclass{article}
\usepackage{spconf,amsmath,graphicx,cite,subfigure}

\title{SNR-ADAPTIVE DEEP JOINT SOURCE-CHANNEL CODING \\FOR WIRELESS IMAGE TRANSMISSION}
\name{Mingze Ding$^{\star}$ \qquad Jiahui Li$^{\dagger}$ \qquad Mengyao Ma$^{\dagger}$ \qquad Xiaopeng Fan$^{\star}$}
\address{$^{\star}$ Harbin Institute of Technology, School of Computer Science $\&$ Technology \\
	$^{\dagger}$ Wireless Technology Lab, Huawei, Shenzhen 518129, China
}
\begin{document}
%
\maketitle
\begin{abstract}
Considering the problem of joint source-channel coding (JSCC) for multi-user transmission of images over noisy channels, an autoencoder-based novel deep joint source-channel coding scheme is proposed in this paper. In the proposed JSCC scheme, the decoder can estimate the signal-to-noise ratio (SNR) and use it to adaptively decode the transmitted image. Experiments demonstrate that the proposed scheme achieves impressive results in adaptability for different SNRs and is robust to the noise in the SNR estimation of the decoder. To the best of our knowledge, this is the first deep JSCC scheme that focuses on the adaptability for different SNRs and can be applied to multi-user scenarios.
\end{abstract}
\section{Introduction}
\label{sec:intro}

According to Shannon's separation theorem \cite{C1}, it is known that under the premise of unlimited time delay and complexity, separate optimization of source coding and channel coding can also achieve the optimal performance. Therefore, most modern communication systems apply a two-step encoding process: First use source coding algorithm to compress source data to remove its redundancy. Then transmit the compressed bitstream over the channel by using a channel coding algorithm. Moreover, current communication systems employ highly efficient source coding algorithms (e.g., JPEG, JPEG2000, BPG) and near-optimal channel codes (e.g., LDPC, Turbo, polar codes) to approach the theoretical optimality. All the source and channel coding algorithms mentioned above have undergone long-term development and are quite effective.

However, in practice, Shannon's hypothesis is not applicable in many cases. Even under these assumptions that Shannon mentioned, the separate approach breaks down in multi-user scenarios \cite{C3,C4}, or non-ergodic source or channel distributions \cite{vembu1995source,gunduz2008joint}. Moreover, in some applications joint source-channel coding (JSCC) scheme is known to be better than the separate approach \cite{C2}.

In this paper, we consider the design of practical JSCC scheme. The proposed JSCC scheme can be adaptive to different SNRs and able to be applied to multi-user scenarios (The SNRs of the users might be different) by using pilot signal as an additional input to the decoder. In actual wireless transimission, the transmitter can send pilot signal known to the receiver. According to the pilot signal, the receiver can estimate the SNR to assist the decoding process.

This work is mainly inspired by recent work on deep JSCC and recent success of deep neural networks (DNNs) (in particular, the autoencoder architectures \cite{bengio2009learning,goodfellow2016deep}). There are many examples of designs about JSCC using autoencoder architectures, e,g.,  \cite{rongwei2003joint,o2016learning,o2017introduction,kim2018communication,nachmani2018deep,caciularu2018blind,o2017deep}. The first work that used neural networks to solve the problem of JSCC is \cite{rongwei2003joint}, where simple neural network architectures were used as encoder and decoder for Gauss-Markov sources over additive white gaussian noise (AWGN) channel. And
in \cite{farsad2018deep}, they considered the problem of JSCC of structured data such as natural language. In \cite{choi2019neural}, they proposed an deep JSCC model which was mainly inspired by low-variance gradient estimation for variational learning of discrete latent variable models. Kurka $et$ $al$. presented several deep JSCC schemes, among which \cite{bourtsoulatze2019deep} provided graceful degradation with the SNR, \cite{kurka2020deep} fully exploited channel output feedback, \cite{kurka2019successive} achieved successive refinement of images. This presented work is most relevant to theirs, especially with \cite{bourtsoulatze2019deep}. The reference \cite{bourtsoulatze2019deep} is their first contribution and this current paper will take it as the baseline.

To the best of our knowledge, this is the first work about deep JSCC scheme that can be adaptive to different SNRs and able to be applied to multi-user scenarios. Although in \cite{bourtsoulatze2019deep}, they presented graceful degradation with the SNR, their results are not quite satisfactory. We conduct sufficient experiments to verify the performance of our proposed SNR-adaptive deep JSCC scheme and compare it with \cite{bourtsoulatze2019deep}, which is state-of-the-art. It can be seen that, the proposed method can better adapt to the changes of the SNR and it is robust to noisy estimation of the SNR.

The rest of the paper is organized as follows. In Section \ref{sec:format}, we introduce the system model and general encoding and decoding process. The proposed model architecture is introduced in Section \ref{sec:pagestyle}. Section \ref{sec:typestyle} presents the evaluation of the SNR-adaptability of the proposed deep JSCC scheme, and its comparison with the baseline. Moreover, Section \ref{sec:typestyle} includes analysis on robustness of the proposed method to noisy SNR estimation. Finally, the paper is concluded in Section \ref{sec:print}.

\section{SNR-Adaptive JSCC process}
\label{sec:format}
\begin{figure}[htb]
	\begin{minipage}[b]{1.0\linewidth}
		\centering
		\centerline{\includegraphics[width=8.5cm]{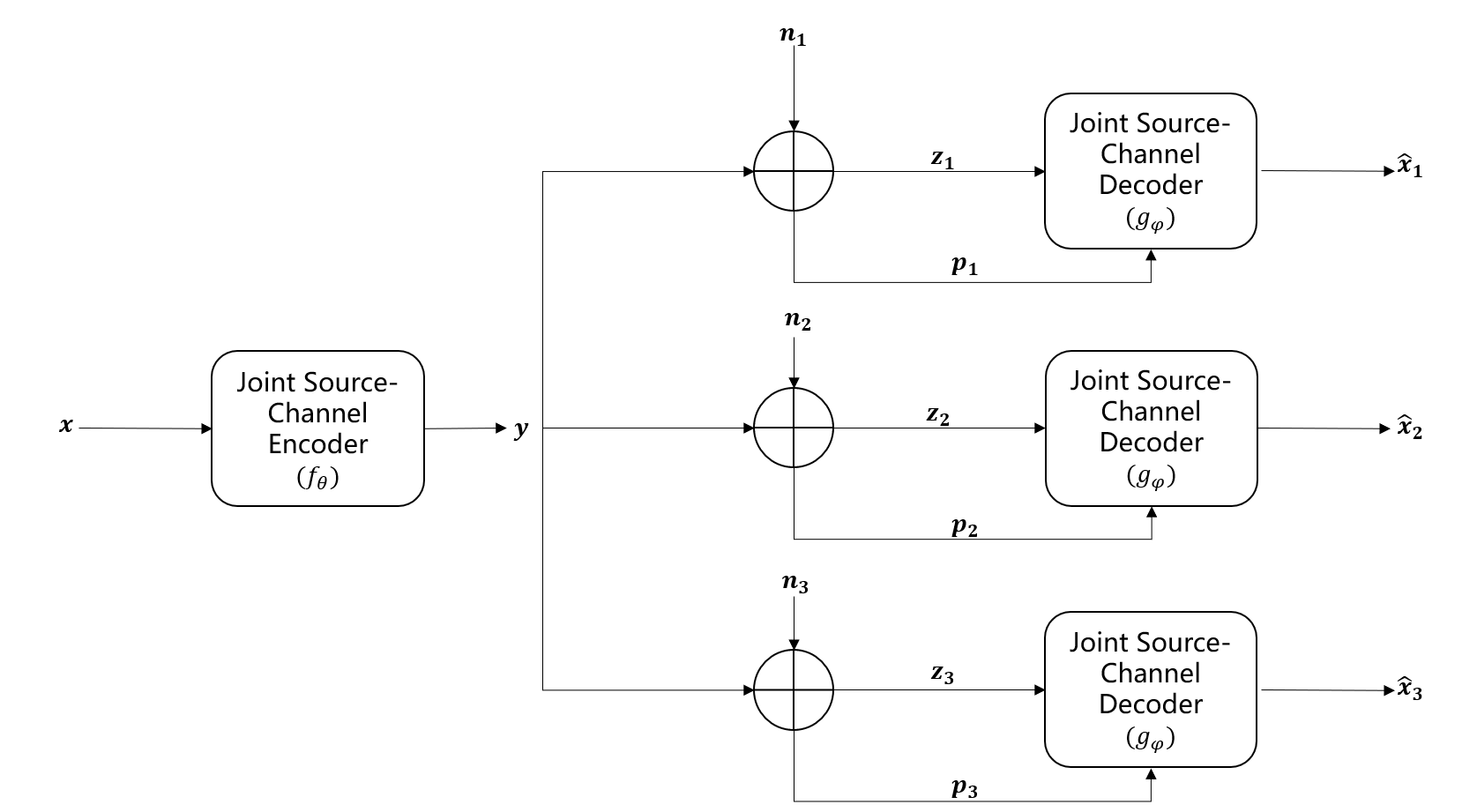}}
	\end{minipage}
	\caption{Multi-user wireless image transmission adaptive to different SNRs. Take the example of facing three users at the same time. The transmitter maps $x$ to $y$. Then $y$ is transimitted over different noisy channels. Finally, the output of the channel $z_{i}$ is fed to the receiver of user $i$. At the same time, the pilot signal $p_{i}$ is also transmitted to the receiver. And the decoder can estimate the SNR by using the received pilot signal.}
	\label{img1}
\end{figure}

Fig. \ref{img1} presents the multi-user wireless image transmission which is adaptive to the SNR. Each user has a different channel but the same decoder. First, the encoder maps $n$-dimensional image $x \in$ $R^{n}$ to a $k$-dimensional vector of complex numbers $y \in$ $C^{k}$. Then apply an average power constraint to $y$ and $y$ satisfies $\frac{1}{k}$$E[y^{*}y]$ $\leq$ $P$. $y^{*}$ is the conjugate transpose of $y$. $P$ is the average transmit power constraint \cite{bourtsoulatze2019deep}. $k/n$ is defined as the bandwidth ratio. And the encoder function $f_{\theta} : R^{n} \rightarrow C^{k}$ is parameterized using a CNN with parameters $\theta$. After encoding, $y$ is transimitted over an AWGN channel and the channel output $z_{i} = y + n_{i}$, where $n_{i}$ is the independent and identically distributed circularly symmetric complex Gaussian noise component with zero mean and variance $\sigma^{2}_{i}$. Based on the pilot signal, the decoder can estimate the average SNR and serve it as auxiliary decoding information. SNR = $10log_{10}\frac{P}{\sigma^{2}_{i}}$. Without loss of generality, we assume that $P = 1$, in following experiments. And the SNR can be changed by adjusting $\sigma^{2}_{i}$. In subsequent experiments, we actually use $\sigma^{2}_{i}$ obtained by the SNR. Finally, the decoder maps $\sigma^{2}_{i}$ and $z_{i}$ to the reconstructed image $\hat{x}_{i} \in R^{n}$. And the decoding function $g_{\varphi} : C^{k} \rightarrow R^{n}$ is parameterized using a CNN with parameters $\varphi$.
\section{proposed model Architecture}
\label{sec:pagestyle}
\begin{figure}[htb]
	\begin{minipage}[b]{1.0\linewidth}
		\centering
		\subfigure[SNR-Adaptive Model]{
		\label{img2a}
		\centerline{\includegraphics[width=8.5cm]{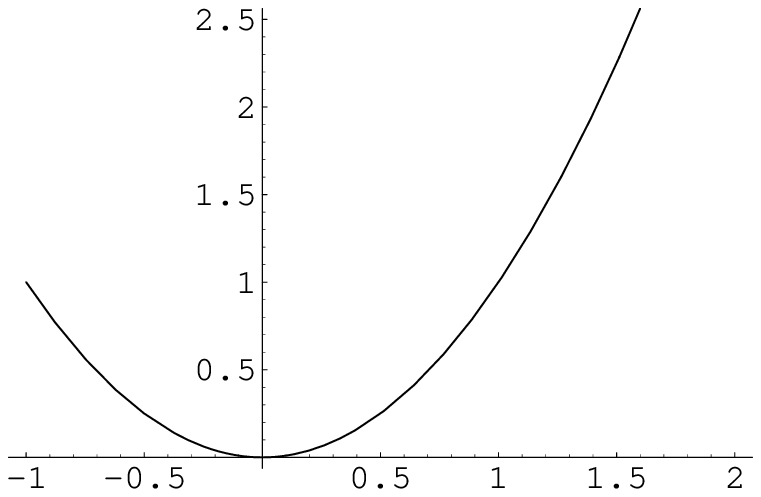}}
		}
	\end{minipage}
	\begin{minipage}[b]{1.0\linewidth}
		\centering
		\subfigure[DM : Denoising Module]{
		\label{img2b}
		\centerline{\includegraphics[width=8.5cm]{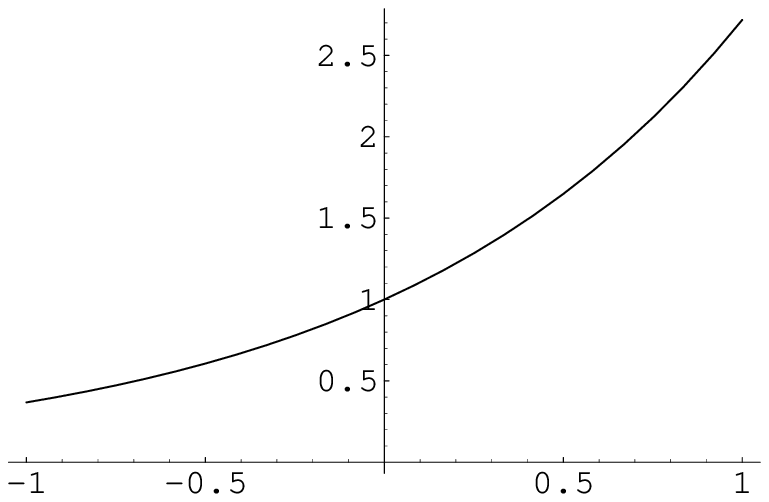}}
		}
	\end{minipage}
	\caption{(a) : The SNR-Adaptive Model. (b) : The architecture of the denoising module (DM).}
	$\bigoplus$ : Add corresponding elements.
	\label{img2}
\end{figure}

The proposed model is mainly based on autoencoder. From Fig. \ref{img2a}, it can be seen that, the encoder consists of five convolutional layers and uses PReLU \cite{he2015delving} activation functions. The notation $K*F*F/S$ denotes a convolutional layer with $K$ filters of spatial extent (or size) $F$ and stride $S$. In the subsequent experimental results, $n$ remains unchanged in encoding. By adjusting the number of filters (C) in the last convolutional layer of the encoder, $k$ can be changed, leading to the change in the bandwidth ratio $k/n$. The decoder consists of deconvolutional layers and also uses PReLU and Sigmoid activation functions. The most important thing is that there are two improvements to the decoder.
\subsection{SNR-Adaptive Decoder}
\label{ssec:subhead3}
The SNR-adaptive decoder has two inputs, one is pilot signal ($p_{i}$) and the other is the channel output ($z_{i}$) obtained by transmitting the encoder output through the noise channel. The decoder uses the pilot signal sent by the transmitter to estimate the SNR, which will be used to assist the decoding process. In subsequent experiments, the estimation of the SNR at the decoder can be divided into two types: with noise and without 
noise. The two inputs are added together (element-wise addition) after passing through a convolutional layer as the input for subsequent operations. In order to achieve the above operation, we expand the single SNR estimated by the decoder to a SNR map, which has the same dimension as $z_{i}$. The value of each element in the SNR map is the value of the channel noise variance estimated by the decoder.
\subsection{Denoising Module}
\label{ssec:subhead4}
In order to better adapt to the changes of the SNR to obtain preferable reconstruction quality. Some improvements are made. We add a Denoising Module (DM) and two short-circuit connections. To some extent, benefiting from the estimated the SNR, DM can learn SNR-adaptive denoising. And short-circuit connection can speed up the convergence of the model. The specific DM can be seen in Fig. \ref{img2b}. The module consists of two branches, one of which is composed of convolutional layer, PReLU activation function and batch normalization (BN) layer \cite{ioffe2015batch} and the other is to replace the convolutional layer in the first branch with dilated convolution (D-Conv) \cite{wei2018revisiting,tian2020image}. The notation $K*F*F/S;D$ denotes a dilated convolutional layer with $K$ filters of spatial extent (or size) $F$ stride $S$ and dilation rate $D$. Employing two branches can reduce the depth of the model to simplify the training. Besides, thanks to different convolutional networks in the two branches, more features can be extracted to achieve better denoising effect. And these two branches are both residual blocks \cite{he2016deep}, the mean of their output is used as the final result and then input it to the subsequent network. The residual network can speed up training and improve the performance of the model to some extent.

\section{EXPERIMENTAL RESULTS}
\label{sec:typestyle}
The above models are implemented in Tensorflow and optimized using the Adam algorithm \cite{kingma2014adam}. The loss function is the average mean squared error (MSE) between the original input image $x$ and reconstruction $\hat{x}$ output from the decoder, defined as:
\begin{equation}\label{eq2}
\mathcal{L} = \frac{1}{N}\sum_{i=1}^{N}d(x_{i}, \hat{x}_{i})
\end{equation}
where $d(x, \hat{x}) = \frac{1}{n}\Vert x-\hat{x}\Vert^{2}$. PSNR is used to evaluate the performance of the proposed method. The PSNR metric measures the ratio between the maximum possible power of the signal and the power of the noise that corrupts the signal. We evaluate the performance of our model on the CIFAR-10 image dataset \cite{krizhevsky2009learning}, which has a training set of 50000 32*32 images and a testing set of 10000 images. Before conducting the experiment, the dataset is preprocessed first. The images in the dataset are 24-bit RGB images. Normalize them and convert the pixel value range from [0,255] to [0,1]. During model training, the learning rate is initially set to $10^{-4}$ and the batch size is 64. Reduce the learning rate to 0.9 times of the original every 10 epochs. Then continue training the model until the test performance no longer shows any improvement. In addition, we first set a training SNR list : [0, 5, 10, 15, 20, 25] (dB) and a testing SNR list : [0, 5, 10, 15, 20] (dB). During training, a SNR is randomly selected from 
\begin{figure}[htb]
	
	\begin{minipage}[b]{1.0\linewidth}
		\centering
		\subfigure[]{
			\label{img3a}
			\centerline{\includegraphics[width=8.5cm]{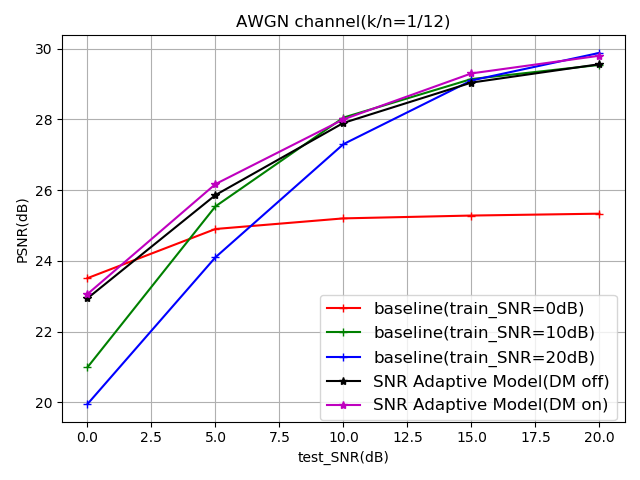}}}
	\end{minipage}
	\begin{minipage}[b]{1.0\linewidth}
		\centering
		\subfigure[]{
			\label{img3b}
			\centerline{\includegraphics[width=8.5cm]{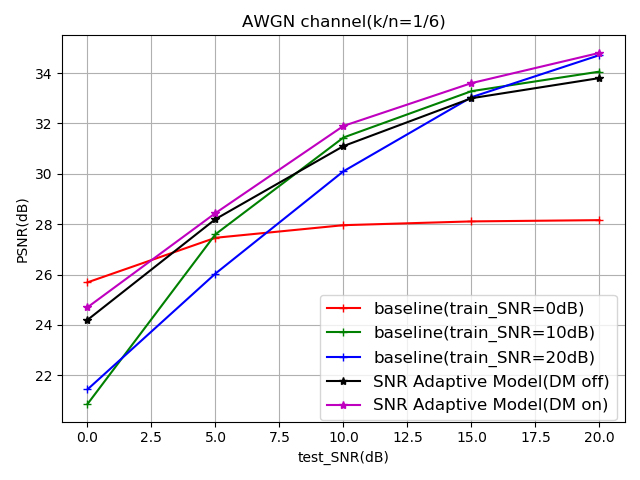}}}
	\end{minipage}
	\caption{Performance comparison for differen test SNRs over an AWGN channel. (a) for bandwidth ratio = 1/12, (b) for bandwidth ratio = 1/6.}
	\label{img3}
\end{figure}
this traning SNR list in each iteration as the SNR in the transmission process. During the test, the SNR in the testing SNR list is used in the transmission process in turn.
\subsection{Adaptability for the SNR}
\label{ssec:subhead1}
First, consider the situation where there is no noise in the estimation of the SNR at the decoder. We compare the performance of the presented SNR-Adaptive Model and the baseline model which is presented in [18]. The training method of the proposed model is as described above. Differently, the baseline model can only be trained by one SNR at a time. Performance of their model in three training SNRs (0, 10 and 20 dB) is presented. In this part, there are two groups of experiments with different bandwidth ratios, which are set 1/12 and 1/6 respectively. From Fig. \ref{img3}, it can be observed that when the testing SNR is lower than the training one, performance of the baseline decreases rapidly as the SNR decreases. And the greater the difference in SNRs for testing and training, the faster the performance degradation. On the contrary, when the training SNR is lower than the testing one, the baseline’s performance improves as the SNR increases. However, the greater the difference in SNRs for testing and training, the slower the performance improvement. In contrast, the performance of the proposed model will also decrease as the testing SNR goes down. But the degradation will not be so severe. And when the testing SNR is lower than baseline’s training SNR, the presented model is superior to the baseline. Besides, when testing SNR is greater than or equal to baseline’s training SNR, the performance of the proposed model with DM off ( When the DM is off, the number of filters in the convolutional layers will also be reduced by 16) is inferior, however, only by a slight difference. As can be seen from Fig. \ref{img2a}, the complexity of the SNR-Adaptive Model does not show great change when the DM is off. Nonetheless, due to the increase of pilot signal as input, the function that decoder needs to learn becomes more complicated. The decoding function changes from the original unary function ($g_{\psi}(z)$) to the binary function ($g_{\psi}(z,p)$). Therefore, when the testing SNR is greater than or equal to the baseline's training SNR, proposed Model with DM off will only be slightly worse. In sharp contrast, the proposed model with DM on not only outperforms model with DM off, but also exceeds the baseline in almost all testing SNRs. So we can say that the
presented approach has better adaptability to the SNR as well as considerable potential. And from this it can be seen that the proposed model still has room for improvement in performance. In the future work, we will strive to improve the performance of the present model.
\subsection{Robustness to Noisy SNR Estimation}
\label{ssec:subhead2}
\begin{figure}[htb]
	
	\begin{minipage}[b]{1.0\linewidth}
		\centering
		\subfigure[]{
			\label{img4a}
			\centerline{\includegraphics[width=8.5cm]{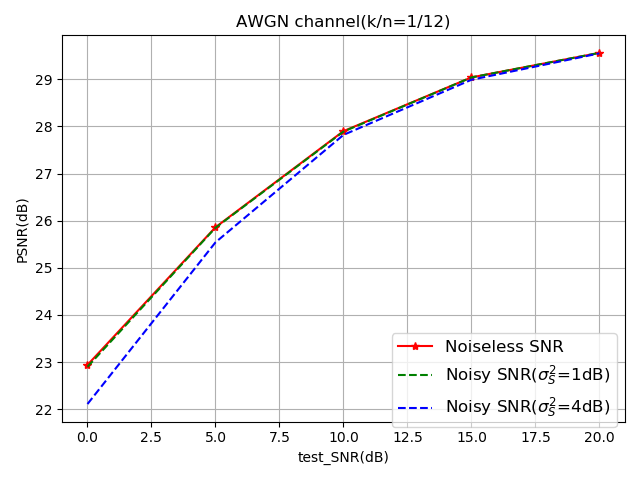}}}
	\end{minipage}
	\begin{minipage}[b]{1.0\linewidth}
		\centering
		\subfigure[]{
			\label{img4b}
			\centerline{\includegraphics[width=8.5cm]{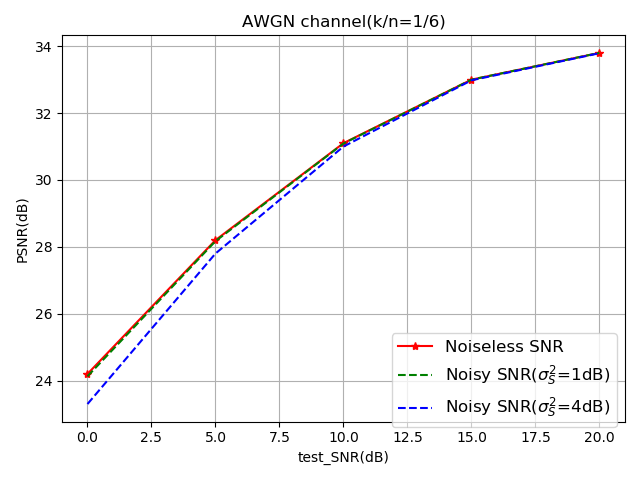}}}
	\end{minipage}
	\caption{Model performance when the decoding end has noise in the SNR estimation. (a) for bandwidth ratio = 1/12, (b) for bandwidth ratio = 1/6.}
	\label{img4}
\end{figure}

In this section, we analyze the robustness of the proposed method when the decoder has noise in the SNR estimation, i.e., $\sigma^{2}_{S}>0$. We first assume that the true value of SNR is $S$. And the estimated SNR is $\hat{S}$, $\hat{S} = S + E$. $E$ obeys a Gaussian distribution with a mean of 0 (dB) and a variance of 1 or 4. Then solve the channel noise variance $\sigma^{2}_{i}$ from the noisy SNR. Since this part of the experiment is to verify the robustness of the SNR-Adaptive Model to noisy SNR estimation, no longer need to compare the baseline with the proposed model with DM on. As can be seen from the Fig. \ref{img4}, the proposed method has the robustness to noisy estimation of the SNR. When the variance of the noise in the SNR estimation is 1 (dB), the performance of the model almost not decreases (the red line and the green line almost completely coincide.). And when the variance of the noise in the SNR estimation is 4 (dB), performance only drops slightly when the testing channel noise is strong.
\section{conclusion}
\label{sec:print}
In this paper, a novel SNR-adaptive deep JSCC scheme is proposed for multi-user wireless image transmission.  
In this scheme, the decoder estimates the SNR from the pilot signal to assist the decoding process and achieve SNR-adaptability. To the best of our knowledge, this is the first method that focuses on the adaptability for different SNRs and can be applied to multi-user scenarios. We present the advantages of the proposed method by comparing it with state-of-the-art. It can be seen from the experiment results that the SNR-adaptability of the method proposed is better than that of the baseline. Since the proposed model is adaptive to the SNR, it has the potential to be applied to multi-user scenarios. What's more, we explored the robustness of our proposed model to noisy estimation of the SNR at the decoder. The results show that even if there exist noise in the estimations of the SNR at the decoding end, the performance of the proposed model only drops slightly even when the noise is relatively strong. 


\vfill\pagebreak


\bibliographystyle{IEEEbib}
\bibliography{refs}

\end{document}